\documentclass[12pt]{article}

\usepackage{arxiv}

\usepackage[utf8]{inputenc} 
\usepackage[T1]{fontenc}    
\usepackage{hyperref}       
\usepackage{url}            
\usepackage{booktabs}       
\usepackage{amsfonts}       
\usepackage{nicefrac}       
\usepackage{microtype}      
\usepackage{lipsum}
\usepackage{amsmath}

 \usepackage{natbib}
 \usepackage{graphicx}
\usepackage{caption}
\author{
  Behrouz ~Behdani\thanks{Corresponding Author} \\
  Department of Chemical and Biomolecular Engineering\\
  Vanderbilt University\\
  Nashville, TN 37212 \\
  \texttt{behrouz.behdani@vanderbilt.edu} \\
   \And
 Hayder ~Alhameedi \\
  Department of Chemical Engineering\\
  Missouri University of Science and Technology\\
  Rolla, MO 65401 \\
  \texttt{haa6hb@mst.edu} \\
}

\begin{document}

\title{\Large \bf Computational Fluid Dynamics Study of Taylor Flow in  Microreactors: Investigating the Effect of Surface Tension and Contact Angle on the Heat and Mass Transfer}
\maketitle
\begin{abstract}
Microfluidics technology offers  high efficiency of heat and mass transfer and  low safety hazards compared to conventional  multiphase processes. The multiphase flow in the microchannels is  usually characterized as Taylor flow that includes elongated  microbubbles , separated by liquid slugs. In the current study, we employ OpenFOAM CFD package to investigate the effect of the surface tension and the contact angle on the interfacial surface area and transport yield  in microscale systems. The results show that contact angle can significantly affect the  surface area and the bubble size while the surface tension does not change these parameters in the system. Moreover, in lower contact angle, the flow may turn into bubbly flow, affecting $K_L$, mass transfer coefficient in the liquid phase. Findings of the current study can improve the mass transfer coefficient in microreactors while avoiding thermal runaways, hot spots  and other safety issues of conventional reactors.
\end{abstract}


\section{Introduction}
Recently, microfluidics has been used in various applications involving heat and mass transfer in microscale systems ( \cite{Behdani2018} \cite{Monjezi2017a} ).  One of the growing applications is to use microfluidics in microreactors. Conventional multiphase reactors like batch, semi batch and multi-stage column reactors have low efficiency in large scale which originates from the low mass transfer rate due to the small interfacial surface area. Moreover, inadequate heat transfer rate may cause thermal runaway, hotspots and other safety hazards. Microreactor technology is a tool to resolve these issues. In these continuous systems, due to the high surface to the volume ratio, mass transfer is improved, and high throughput of product is within reach. On the safety aspect, low liquid holdup in the microreactor reduces the risk of working with hazardous materials. Moreover, higher surface area enhances the heat transfer and prevent process hazards like thermal runaways and hotspots in the reactor( \cite{Gemoets2016} \cite{Siddiquee2016} ).\\
In microreactors, the flow regime depends on the physical properties and flow rate of the liquid and gas phases and geometry of the microreactor. The flow regime can have different patterns including Taylor (slug) flow, bubble flow, annular flow, and Spray flow among which the most common is Taylor flow ( \cite{Kashid2007}\cite{Sobieszuk2010} ). Taylor flow is determined by the presence of a gas bubble moving through a microchannel and separated by liquid slugs. There are two mechanisms for Mass transfer in Taylor flow regimes: convection within the liquid phase and diffusion between the liquid slug and the bubble ( \cite{Zhang2018} ). The convection in the liquid slug is due to the recirculation in the slug which originates from the shear stress between the liquid and micro channel wall. The diffusion is caused by the concentration gradient between gas and liquid phase. For convection inside the liquid slug, length of the bubble and liquid slug in the microchannel is significant. It has been shown that if the length of the liquid slug in lower than the diameter of the microchannel, the recirculation does not happen. On the other hand, the diffusion mainly depends on the interfacial area between two phases\cite{Sobieszuk2011}. Therefore, to improve the mass and the heat transfer in microreactors with Taylor flow, increasing the interfacial area and relative length of the bubble and the liquid phases can be influential \cite{Kashid2007a}.\\
Due to the high surface to volume ratio, surface forces become significant in hydrodynamic of microfluid setups ( \cite{Waelchli2006} ). Surface forces depend on the structure of the surface (wettable and non-wettable surface) and surface tension of liquid in the microchannel.  In several studies, thermal or photo active coating or functional groups have been utilized to control the flow regimes in the microchannels. Although these studies show the possibility of changing the interfacial area and relative length of the bubble and liquid phases , to the best of our knowledge, there is no study in the literature which can address the effect of contact angle and surface tensions on these two parameters and more broadly on the flow regime in the microchannel.  In the present study, we  utilize computational fluid dynamics to investigate the effect of surface tension and contact angle in Taylor flow inside the microchannel to understand how they affect the interfacial area and the relative length of the bubble and liquid phases.

\section{Methods}
In this study, Volume of Fluid (VOF) method is employed to investigate the Taylor flow in the microchannel. VOF is an Eulerian approach which is used to simulate the flow dynamics of two or more immiscible fluids. In this approach, a set of continuity and momentum equations as well as equations for tracking the volume fraction of each phase are solved simultaneously. VOF is considered as a conservative, fast and quite simple approach to track the interface of multiphase systems.  Its application for gas-liquid and liquid-liquid multiphase systems has been reported extensively in the literature ( \cite{Deshpande2012} \cite{Li2013} ). 
In microfluidics, the Reynolds number is less than 1 and the flow is considered as laminar flow. In the current study, both phases are considered as Newtonian and incompressible fluids. The continuity and momentum equations are as follows:
\begin{equation}
 \dfrac{\partial \rho}{\partial t} + \nabla .(\rho u ) = 0
\end{equation}
\begin{equation}
 \dfrac{(\partial \rho u)}{\partial t} + \nabla .(\rho u u) = -\nabla p +\mu \nabla^2 u + \rho g
\end{equation}
Here, u, p are the velocity vector and pressure, respectively. $\rho$ and $\mu$, in continuity and momentum equations, are the volume- averaged density and viscosity of both phases, as defined by following equations:
\begin{equation}
\rho = \rho_l \alpha +\rho_g (1-\alpha)
\end{equation}
\begin{equation}
\mu = \mu_l \alpha +\mu_g (1-\alpha)
\end{equation}
Interfacial surface is tracked with an additional advection equation as follows:
\begin{equation}
 \dfrac{\partial \alpha}{\partial t} + \nabla .(\alpha u ) = 0
\end{equation}
Where $\alpha$ is the volume fraction of liquid phase. $\alpha=1$ shows liquid phase while $\alpha =0$ shows the gas phase and $0<\alpha<1$ represents the interface of gas and liquid.\\
The adhesion to the wall is considered by using contact angle boundary condition in the solution. It is assumed that the contact line is independent of flow’s direction and velocity i.e. contact angle is constant.\\
To solve the governing equations, InterFoam solver in the OpenFOAM package is used. OpenFOAM is built based on $C++$ libraries to solve the finite volume equations.  Momentum equation solution in InterFoam is done through a predicted velocity field. The predicted velocity field is then corrected by using PISO (Pressure-Implicit with Splitting of Operators) which gives the pressure and velocity in the next time-step. Available studies in literature show that InterFoam is accurate (quantify this) for phenomena with dominant effect of surface tension ( \cite{Deshpande2012} ). In this solver, surface term is considered as a body force only in the interfacial cells ( \cite{Li2013} ).\\
One issue with VOF is the excessive diffusion in the gas-liquid interface, leading to blurred surfaces. In Interfoam, however, to decrease the diffusion and maintain the sharpness of surfaces, a numerical solution for interface compactness is presented ( \cite{Behdani2020a} ).  For that, equation 5 is changed as follows:
\begin{equation}
 \dfrac{\partial \alpha}{\partial t} + \nabla .(\alpha u ) -\nabla . (\alpha (1-\alpha) U_r ) = 0
\end{equation}
Where $U$ and $U_r$   interface velocity and compression velocity:
\begin{equation}
U = u_l \alpha +u_g (1-\alpha)
\end{equation}
\begin{equation}
U_r = u_g - u_l
\end{equation}

\begin{figure}[tbp]
\centering
\includegraphics[width=1\columnwidth]{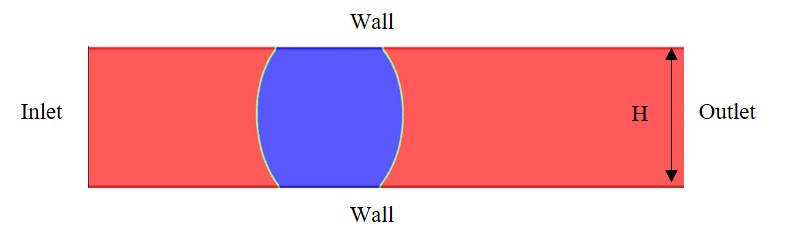}
\caption{. Simulation Geometry: Bubble and liquid slug in the microchannel}
\label{Geometry}
\end{figure}

\begin{figure}[tbp]
\centering
\includegraphics[width=0.6\columnwidth]{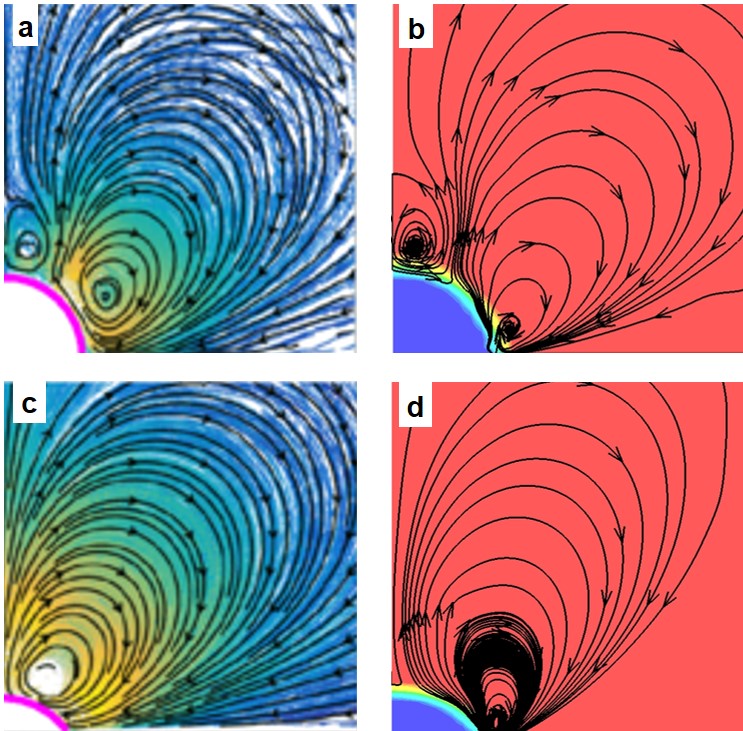}
\caption{ . Model validation : Numerical results are compared with the experimental results for a microbubble streaming system}
\label{Validation}
\end{figure}

$U_r$ term is only present in the interface and disappears in gas and liquid bulk calculations. Ur is calculated in InterFoam as:
\begin{equation}
U_r = n_f min[C_\alpha \mid \Phi/S_f \mid, max(\mid \Phi \mid/ \mid S_f \mid )]
\end{equation}
where, $n_f$, $\Phi$ and $S_f$ are normal flux, face volume flux and surface area of cell. $C_{\alpha}$ is a compression factor which changes from zero to 4. Although higher compression factor gives sharper interface, for values above one it may cause inaccurate results in the simulation.\\
Similar to other finite volume approaches, the timestep is controlled by Courant number (Co) ( \cite{Nekouei2017} ). Co=1 shows that fluid element is moved for one grid size in one time step. To maintain stability of the solution, Co number should be lower than one and has the maximum value of 0.3 in current study.\\
Regarding discretization, for time derivative term, first order implicit Euler is used.  Divergence term in advection equation for tracking surface is discretized by the second-order Gaussian linear integration scheme with van Leer limiter. For discretization of other gradient and divergence term, second order Gaussian  linear integration scheme is utilized. To meet convergence criteria, residual  for continuity should be lower than 1e-6  and for the pressure correction , it should be below 1e-7.\\
 Figure \ref{Geometry} shows the geometry of microchannel. For outlet, the zero gradient boundary condition is maintained. On the walls, no slip boundary condition for velocity and contact angle boundary condition for liquid fraction are enforced. The computational domain includes 44000 hexahedral mesh. Our grid tests showed that meshed finer than 30,000 grids do not give significant differences in the results. To enforce the slug flow, the simulation has two steps. Initially, by setting liquid fraction to zero in the inlet, the gas is injected in the microchannel and then by changing the liquid fraction to 1 , the liquid is injected in the microchannel.\\
The model was validated in our previous work on the microbubble streaming ( \cite{Behdani2020} ). In microbubble steaming, a microbubble is protruded into the main channel. When the microbubble is formed, an oscillation is applied to the microbubble to generate a secondary flow in the microchannel. The flow patterns in microbubble streaming are important in different applications like microparticle separation and micromixing. Numerical results were compared to experimental results by Volk and Kahler ( \cite{Volk2018} ).  Figure \ref{Validation} shows the comparison numerical simulation and experimental results.
\section{Results and discussions}
\subsection{The effect of  Contact angle on slug length}
\begin{figure}[tbp]
\centering
\includegraphics[width=1\columnwidth]{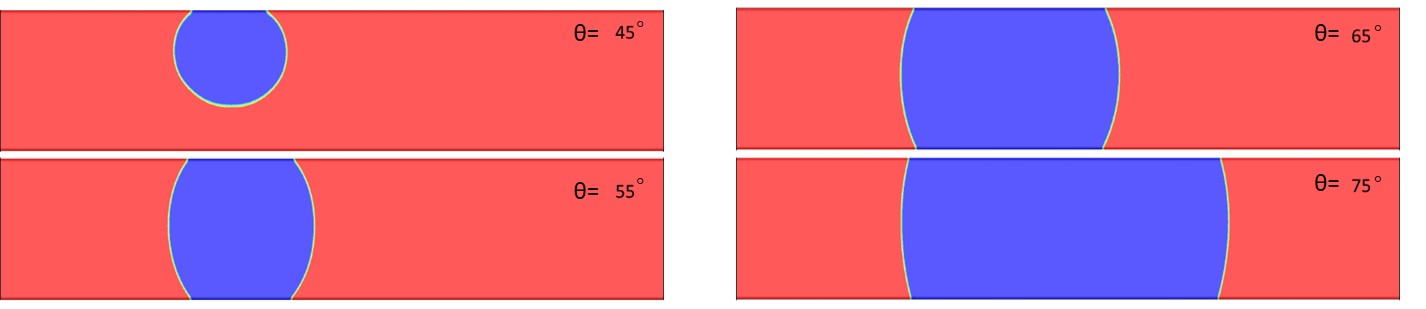}
\caption{. Bubble and liquid profile in different contact angles. As the contact ange increases, the size of the bubble increases.}
\label{Alpha}
\end{figure}

 Figure \ref{Alpha} shows the effect of  contact angle on the length  of bubble in the same injection rate of liquid. Dimensions of the bubble are highly dependent on the  contact angle. As contact angle increases, length of gas phase increases. Moreover, the ability of the bubble to take all the cross section of  microchannel depends on the contact angle. At contact angles , as low as 45 degrees, bubble  is not  able to cover the total cross section of microchannel while with the increase of the contact angle to 55 degrees, the bubble will cover all the cross section.\\
 \begin{figure}[tbp]
\centering
\includegraphics[width=0.5\columnwidth]{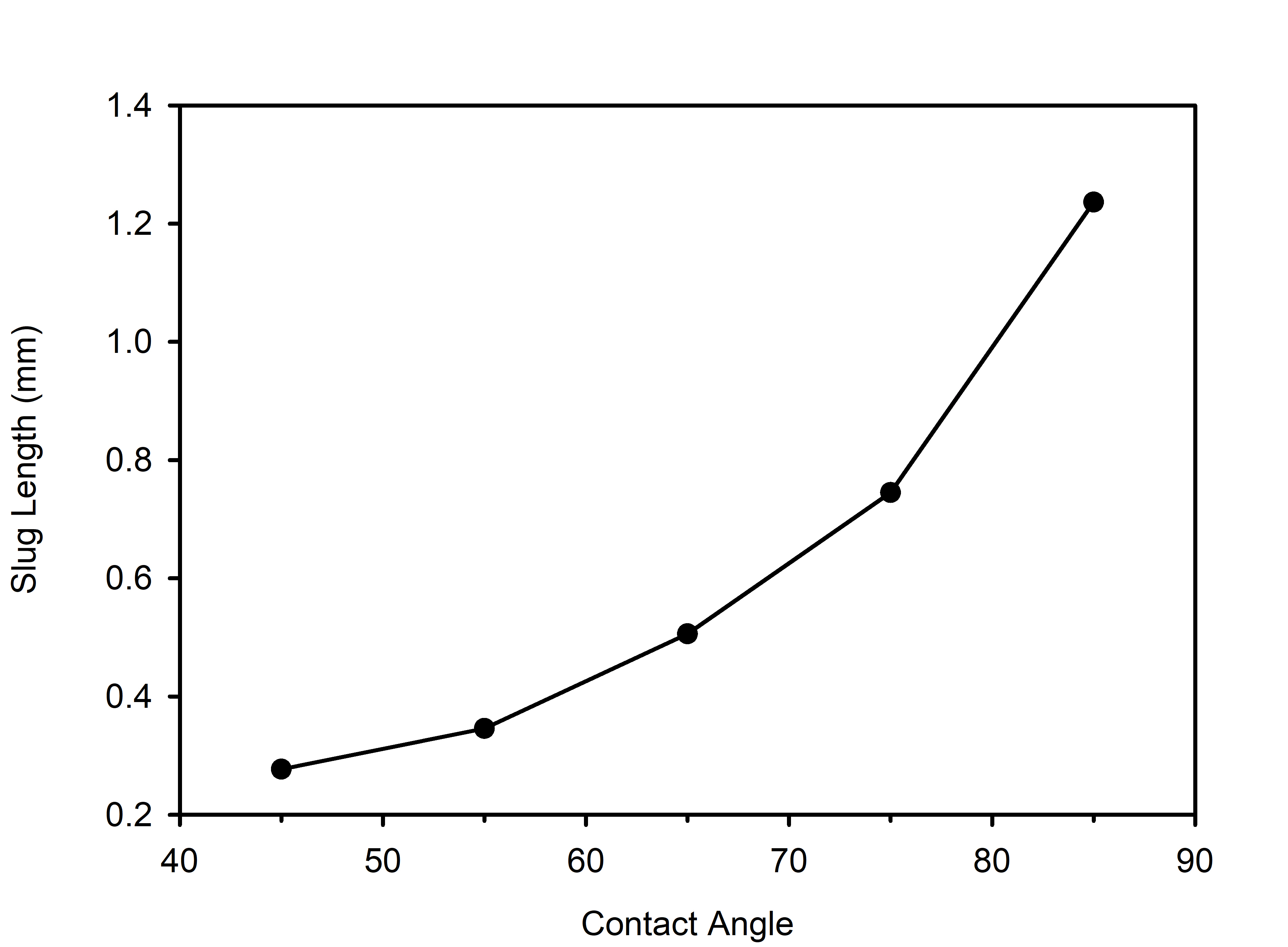}
\caption{. Effect of the contact angle on the area of the microbubble. As the contanct angle increase, the length of the bubble changes significantly.}
\label{Slug Length}
\end{figure}
Figure \ref{Slug Length} shows the effect of contact angle on the length of slug in the microchannel. As the contact angle increases, for the same outlet pressure and inlet flow condition, the length of slug flow increases. As the contact angle increases, the effect of surface tension becomes more and more negligible and the length of the  bubble phase increases. The same trend was observed in previous experimental work for droplet on the surface ( \cite{Drelich1997} ). Moreover, it has been reported in the literature that cos $\theta$ increases with 1/r  where $\theta$ and r are contact angle and curvature radius respectively. In order to obtain curvature radius,  surface is fitted on the interface of liquid and bubble phase. The surface is reconstructed through $\alpha$-shape technique which bounds a thin 3 dimensional layer on the gas-liquid interface.\\
Mass transfers in the microchannel with slug flow is limited by the mass transfer in the liquid phase. In order to improve the mass transfer rate, $k_L$ , mass transfer in the liquid phase, should increase and it is made possible through increasing the recirculation in the liquid phase.  Recirculation in the liquid phase depends on the size of the liquid slug and is absent if the liquid slug size is smaller than the diameter of the microchannel. By increasing, the contact angle, the relative size of the bubble size to liquid slug increases, so it can be understood that moving through hydrophobic liquid does not help mass transfer in the system.
\subsection{Effect of Contact Angle on the Mass Transfer Area}
\begin{figure}[tbp]
\centering
\includegraphics[width=1\columnwidth]{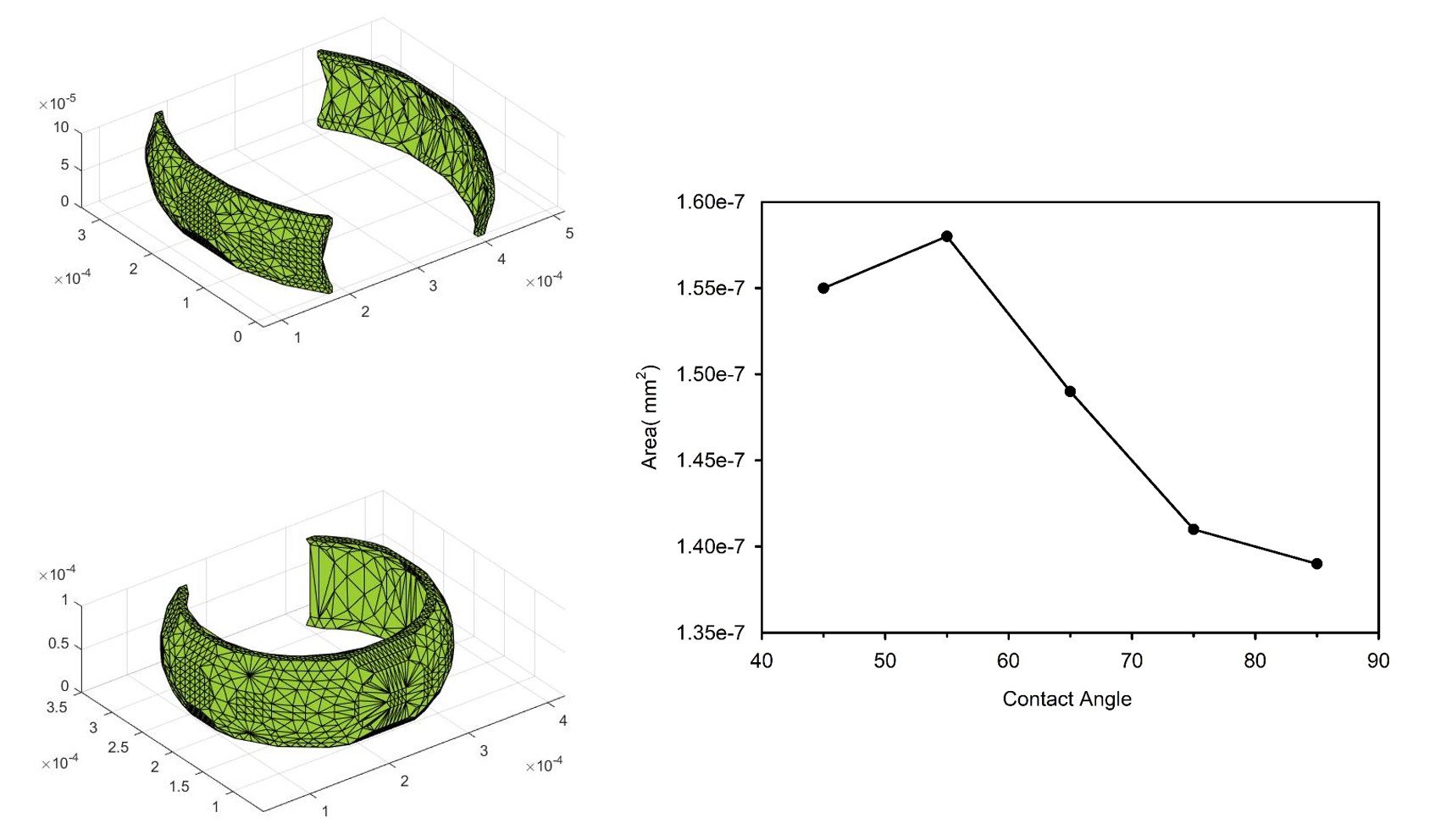}
\caption{. Effect of the contact angle on the area of the microbubble. With the increase in the contact angle, at first the surface area increases since the bubble starts to fill whole cross section of the microchannel. However, with the further increase of the contact angle , the surrface area  decreases.}
\label{Area}
\end{figure}

The other significant parameter which defines the efficacy of mass transfer in slug flow system is  the interface of gas and liquid phases. Figure \ref{Area} shows the surface area of reconstructed interface at different contact angles. If the bubble covers all the cross section of channel, with the increase contact angle, curvature of the interface decreases so it should be expected that by increasing contact angle the surface area decreases. However, by decreasing the contact angle to 45 degrees, bubble is not able to cover whole cross section of the microchannel  leading to decrease in to decrease in the surface area. Moreover, the slug flow regime converts to the bubble flow regime.
\subsection{Effect of Surface Tension on the surface area}
\begin{figure}[tbp]
\centering
\includegraphics[width=0.5\columnwidth]{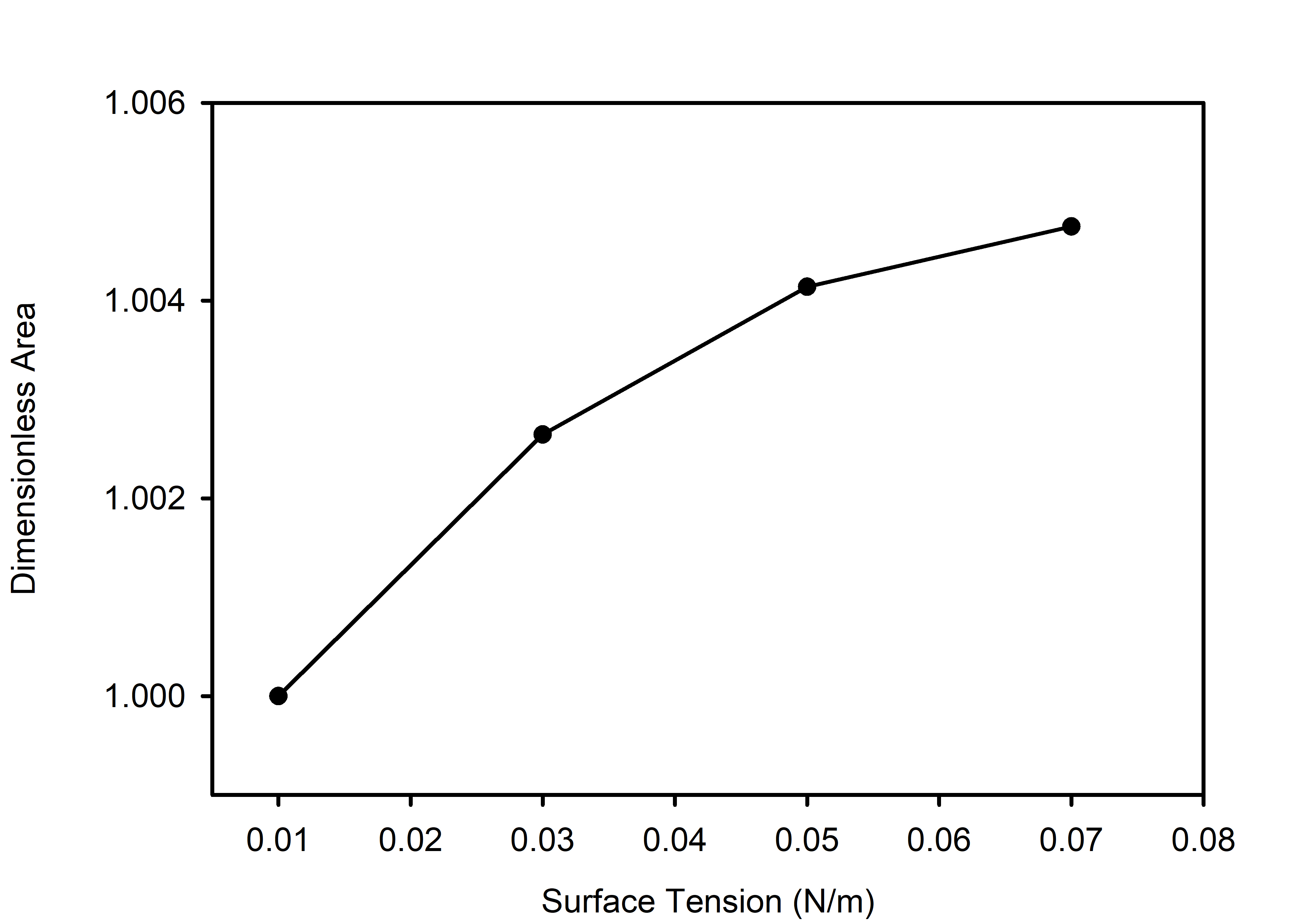}
\caption{. Effect of the surface tension on the surface area of the microbubble. The dimensionless surface area is defined based on the area for the surface tension 0f 0.01 $N/m$}
\label{Surface Tension}
\end{figure}

In different applications, different solvent and liquids are utilized  with different surface tensions. In our previous work, we established that in microbubble streaming with the increase of  surface tension, maximum protrusion of the bubble before bursting into the main channel increases. It shows that with the increase of surface tension , the curve of the bubble increases. Therefore, it should be expected with the increase of the liquid surface tension, the interface surface area increases. Figure \ref{Surface Tension} shows the surface area with the increase of surface tension. With the increase of surface tension, the  interface area does not change very significantly compared to microbubble streaming application. It should also be noted that if there are about 1000 bubble in the setup, then the change of surface area will then affect the surface area for mass and heat transfer in the system. As discussed before, it can be concluded that by increasing  the surface tension , increase of mass transfer in expected.
\section{Conclusion}
The high surface to volume ratio in the microchannel flows helps improve heat and mass transfer in the microreactors. In the current study, to optimize the process in the microscale, flow patterns and the effect of surface related forces are investigated.\\
Increase in the contact angle causes the  non linear increase in the bubble length. In a constant length of the microchannel, the increase in the bubble length means liquid slug possess smaller size which leads to decrease  in the mass transfer in the microchannel.To improve heat and mass transfer between phases, the surface area between phases should increase. With the increase of the contact angle from 45 degrees to 55 degrees, since the flow changes from the bubbly flow to Taylor flow , the surface area increases. However, with further increase the contact angle, the surface area and so heat and mass transfer coefficient decreases. The increase in the surface tension will cause the increase in the surface area, although its effect is not as strong as the effect of the contact angle. \\
The current study investigates isothermal systems and does not consider the property changes due to the temperature. In future studies, nonisothermal system should be studies and the dependence of Nusselt and Sherwood number of the system should be calculated and then compared with the experimental results.






\end{document}